\documentclass[showpacs]{revtex4}
\usepackage{amsmath,amssymb}
\usepackage{natbib}
\usepackage{url}
\usepackage{graphicx}
\usepackage{psfrag}

\usepackage{mathrsfs}
\DeclareSymbolFontAlphabet{\mathrsfs}{rsfs}
\DeclareMathAlphabet{\mathcal}{OMS}{cmsy}{m}{n}

\newcommand{\scri}{\mathrsfs{I}}

\newcommand{\be}{\begin{equation}}
\newcommand{\ee}{\end{equation}}

\def\tt{{\tilde{t}}}
\def\tr{{\tilde{r}}}
\def\tg{{\tilde{g}}}
\def\tq{{\tilde{q}}}

\def\tw{\widetilde{\mathcal{M}}}

\begin{document}


\preprint{AEI-2007-177}
\title{Hyperboloidal foliations and scri-fixing}
\author{An{\i}l Zengino\u{g}lu}
\email{anil@aei.mpg.de}
\affiliation{Max-Planck-Institut f\"ur Gravitationsphysik,
	     Albert-Einstein-Institut,
	     Am M\"uhlenberg~1, D-14476 Golm, Germany}


\begin{abstract}
A useful choice of gauge when including null infinity in the
computational domain is scri-fixing, that is, fixing the spatial
coordinate location of null infinity. This choice allows us to avoid
the introduction of artificial timelike outer boundaries in numerical
calculations. We construct manifestly stationary scri-fixing
coordinates explicitly on Minkowski, Schwarzschild and Kerr
spacetimes.
\end{abstract} 


\maketitle
\section{Motivation}
In numerical studies of the initial value problem for test fields on
asymptotically flat background spacetimes, one typically truncates the
solution domain by introducing an artificial timelike outer boundary
into the spacetime. The solution then is calculated on a finite,
spatially compact domain. The boundary of this domain is
not part of the physical problem. Therefore, one tries
to construct boundary conditions that correspond to transparency of
this artificial outer boundary. In addition, these boundary conditions
are required to form a well posed initial boundary value problem
(IBVP).

In general, it is not possible to construct such boundary
conditions. Spurious reflections occur from the outer boundary even
for the simple case of the flat wave equation on a three-dimensional
ball \cite{Sarbach07}. One therefore tries to minimize the amount of
such spurious reflections in a manner that also ensures the
wellposedness of the IBVP. Such boundary conditions are called
non-reflecting or absorbing. It is, however, very difficult to
construct them when the curvature of the background does not vanish or
when non-linear terms appear in the equations. One needs to account
for backscatter off curvature or self-interaction of the field near
the boundary. A bad choice of boundary data can destroy relevant
features of the solution. It has been shown, for example, that a
certain choice of boundary data, commonly used in numerical
relativity, destroys the polynomial tail behavior of solutions to wave
equations on a Schwarzschild spacetime \cite{Dafermos04}.

A further difficulty is related to the notion of radiation. In the
study of radiative phenomena, one is interested in the behavior of
solutions in the far-field zone. The electromagnetic field on flat
space, for example, admits a global separation into near field terms
and far field terms which simplifies the discussion of radiation
considerably. When the background curvature does not vanish, such a
separation can only be expected in the asymptotic limit to infinity in
null directions. Therefore to study radiation accurately, one needs to
calculate very large portions of spacetime. Using Cauchy-type
foliations and truncating the computational domain, however, does not
allow us to calculate such large portions of spacetime that
astrophysically realistic distances can be modeled in a numerical
calculation.

Additional complications arise when one tries to construct absorbing
boundary conditions for the Einstein equations
\cite{Sarbach04,Rinne06, Buchman07, Friedrich:2005kk}. The treatment
of the asymptotic region in numerical relativity is an important
theoretical and practical problem as it effects the accuracy of the
calculated radiation field \cite{Rinne07, Pazos06}.  With increasing
sensitivity of detectors and computer power, systematic errors in
current numerical simulations due to wave extraction may restrict the
predictable power of numerical relativity and blur the comparison
between observed and calculated waveforms \cite{Boyle:2007ft,
  Hannam:2007ik}.

A clean solution to the above mentioned difficulties is available on
the level of geometry.  The solution is to include null infinity in
the computational domain.  This geometric idea goes back to Penrose
and has been known for a long time \cite{Penrose63, Penrose65}. Its
implementation within the 3+1 approach, however, has been exceptionally
difficult, numerically as well as analytically.  In this article we
discuss in detail the construction of suitable explicit gauges on
asymptotically flat background spacetimes such that the geometric idea
of null infinity can be incorporated in numerical calculations within
the 3+1 approach. While we assume the spacetime to be given, a basic
motivation is to extend the suggested methods to the treatment of the
asymptotic region in numerical solutions to the Einstein equations.

The article is organized as follows: In section \ref{sec:main} we
introduce the basic concepts and explain the idea of scri-fixing. In section
\ref{sec:sphsym} we present an explicit construction of scri-fixing
coordinates in spherical symmetry. This method is applied to
Minkowski spacetime in section \ref{sec:mink} and to
extended Schwarzschild spacetime in section \ref{sec:ss}. Going beyond
spherical symmetry, we present the construction of a scri-fixing gauge
on Kerr spacetime in section \ref{sec:kerr}.
We conclude in section \ref{sec:conc} with a
discussion of scri-fixing and an outlook for future work.

\section{The general idea}\label{sec:main}
If a background spacetime $(\widetilde{\mathcal{M}}, \tg)$ is
asymptotically flat in a suitable sense, one can attach to the
spacetime manifold $\widetilde{\mathcal{M}}$ a boundary in null
directions, $\scri$, such that a certain conformal rescaling of the
physical metric $\tilde{g}$ results in a metric $g=\Omega^2\tilde{g}$
with a smooth extension through that boundary and the conformal factor
$\Omega$ has a specified behavior on that boundary. In particular, we
require $\Omega|_{\widetilde{\mathcal{M}}}>0$,
\mbox{$\Omega|_{\scri}=0$} and \mbox{$d\Omega|_{\scri}\ne0$}. The
extended manifold with boundary is denoted by $\mathcal{M}$. We call
the triple $(\mathcal{M},g,\Omega)$ a conformal extension and say that
the spacetime $(\widetilde{\mathcal{M}}, \tg)$ is weakly
asymptotically simple. All known explicit solutions to the Einstein
equations that may be considered as "physically reasonable" and
asymptotically flat are weakly asymptotically simple \cite{Hawking73}.

By construction, null rays reach $\scri$ for an infinite value of the
affine parameter along them. Therefore we refer to $\scri$ as null
infinity. It turns out, under reasonable conditions, that $\scri$
consists of two parts $\scri^+$ and $\scri^-$, called future and past
null infinity respectively, each with topology $\mathbb{R}\times
S^2$ \cite{Penrose65}. We will mainly focus on future null infinity.

The propagation of massless fields is governed by the null cone
structure of spacetime. It is therefore convenient in studies of
radiative properties of test fields to choose a foliation that is
closely related to outgoing null directions. An option is to choose a
characteristic foliation. This approach has been very successful in
situations where null foliations cover the computational domain in a
regular fashion \cite{Bartnik99, Gomez:1998uj, Gomez:2002ev,
  Husa:2001pk, Winicour05}. The main difficulty with this approach can
be traced back to the fact that characteristic foliations are not
well-behaved in regions of strong dynamical gravitational fields due
to formation of caustics in the bundles of light rays generating the
null hypersurfaces \cite{Friedrich83}.

Spacelike foliations are more flexible than characteristic ones. It
turns out that spacelike surfaces can be constructed that extend
through null infinity. Such surfaces are called \textit{hyperboloidal}
as their asymptotic behavior is similar to standard hyperboloids in
Minkowski spacetime \cite{Friedrich83a}. Instead of approaching
spatial infinity as Cauchy surfaces do, they reach null infinity,
$\scri$, which makes them suitable for radiation extraction. It should
be emphasized that hyperboloidal surfaces extend through
null infinity as spacelike surfaces. It would be misleading to refer
to them as asymptotically null because their tangent vector never
becomes null.  The term asymptotically null may be reserved for
surfaces that become null at or close to null infinity, in which case,
for example, their mean extrinsic curvature would be unbounded.

One can formulate a Cauchy problem based on hyperboloidal surfaces
\cite{Geroch77, Friedrich83a}.  Although the hyperboloidal initial
value problem and its merits have been known for a long time,
numerical studies of test fields based on hyperboloidal foliations are
rare. One major difficulty is related to the choice of gauge. Future null
infinity is an ingoing null surface. This property can be made
manifest in the gauge such that null generators of $\scri^+$
converge. Such coordinates can be useful in numerical studies for
discussing global properties of spacetimes, for example with respect
to the investigation of a point corresponding to timelike infinity
$i^+$ \cite{Huebner01, Zeng06a}. These coordinates, however, do not
seem to be convenient for calculating radiation along $\scri^+$, as
they imply a resolution loss in the physical part of the conformal
extension and require a gauge and boundary treatment in the unphysical
part.  We emphasize that, as also follows from abstract, semi-global
existence results \cite{Friedrich86}, the resolution loss is not an
intrinsic deficiency of conformal compactification, but a consequence
of a bad choice of gauge for the type of problem that one is
interested in.

A suitable class of gauges on hyperboloidal foliations for studying
radiation is given by scri-fixing \cite{Frauendiener98b}. In such a
gauge the spatial coordinate location of null infinity is independent
of the time coordinate so that $\scri^+$ is represented by the same
grid points during time evolution and no resolution loss appears.
This allows us to calculate radiation emitted to future null infinity
in a very accurate way using a simple extraction procedure along an a
priori known grid surface. Further, one can let the numerical outer
boundary coincide with $\scri^+$ so that no boundary treatment on the
continuum level is required \cite{Husa05}. It should be noted that in
a characteristic approach based on a double null gauge, scri-fixing is
automatically satisfied by the coordinates. This seems to be one of
the key advantages in including $\scri^+$ in the computational domain
within the characteristic approach. Within the hyperboloidal approach,
the larger freedom of gauge choices needs to be carefully dealt with
to achieve scri-fixing in the asymptotic domain.

The first explicit construction of a scri-fixing gauge has been given
for Minkowski spacetime \cite{Moncrief00,Husa02b}. Its usefulness has
been demonstrated in studies of dynamical magnetic monopoles on a flat
background \cite{Fodor03,Fodor06}. The explicit constructions
presented in this article should be useful for studies of test fields
on black hole spacetimes \cite{Zenginoglu:2008wc}. They are also
intended to provide examples for testing ideas on treating the
hyperboloidal initial value problem for the Einstein equations as well
as for constructing hyperboloidal initial data \cite{Zeng08b,
  Andersson02a}.

A convenient way to discuss global properties of foliations is to
depict their embedding in Penrose diagrams. We will employ such
diagrams in this article to plot results of calculations. In spherical
symmetry, Penrose diagrams accurately represent the global causal structure
of the spacetime under study (see \cite{Dafermos05} for a rigorous
discussion of Penrose diagrams).

\section{Scri-fixing in spherical symmetry}\label{sec:sphsym}
The main property of a scri-fixing gauge is that the zero set of the
conformal factor in terms of local coordinates is independent of
time. There are two steps involved in the construction of such a
gauge: the choice of a hyperboloidal foliation and a conformal
compactification, that is, the introduction of compactifying spatial
coordinates in combination with a suitable conformal factor. We
present in the following how these steps can be carried out explicitly
for asymptotically flat, spherically symmetric spacetimes.  The
techniques can be easily carried over to spacetimes with less
symmetries as demonstrated in section \ref{sec:kerr}.

\subsection{Hyperboloidal foliations}
Assume that we are given a globally hyperbolic, asymptotically flat,
spherically symmetric spacetime $(\tw,\tg)$ so that the group $SO(3)$
acts non-trivially by isometry on $(\tw,\tg)$. We can introduce on the
quotient space $\widetilde{\mathcal{Q}}=\tw/SO(3)$ a natural function
$\tr:\widetilde{\mathcal{Q}}\to\mathbb{R}$, called the area-radius,
such that the group orbits of points $p\in\widetilde{\mathcal{Q}}$
have area $4\pi \tr(p)^2$.  The metric $\tg$ on $\tw$ can then be
written as $\tg = \tq + \tr^2\,d\sigma^2$, where $d\sigma^2$ is the
standard metric on $S^2$ and $\tq$ is a Lorentzian metric of rank 2
that descends naturally to a metric on $\widetilde{\mathcal{Q}}$. Due
to the Jebsen-Birkhoff theorem \cite{Johansen05}, we can assume that,
beside the $SO(3)$ generators, there is an additional hypersurface
orthogonal Killing vector field on $\tw$ that is timelike in the
exterior \cite{Ehlers06}. We can introduce a time function $\tt$ such
that this Killing vector field is given by $\partial_{\tt}$ and level
sets of $\tt$ are orthogonal to it. In these coordinates, the
spacetime metric reads \be\label{eq:phymet} \tg = \tq_{\tt\tt}\,d\tt^2
+ \tq_{\tr\tr}\,d\tr^2 + \tr^2\,d\sigma^2, \ee where $\tq_{\tt\tt}$
and $\tq_{\tr\tr}$ are functions of $\tr$ only. The time function
$\tt$ is by construction such that the time slices are Cauchy surfaces
and approach spatial infinity in the asymptotic region. We are
interested in null infinity. Therefore, we need to introduce a new
coordinate $t=t(\tt,\tr)$ such that the time slices are hyperboloidal
surfaces.

In stationary spacetimes one may define the notion of \emph{Killing
  observers}. The worldline of a Killing observer corresponds to the
integral curves of the timelike Killing field. These observers are not
inertial if the curvature of the spacetime does not vanish, but they
are natural in the sense that their distance to each other remains
constant.  A Killing observer far away from a source can be regarded
as representing a gravitational wave detector. Therefore, we wish to
keep the natural coordinate representation of the Killing observers so
that the Killing vector field is given by $\partial_t$ in terms of the
new coordinate $t$.  This implies a transformation of the form
\be\label{eq:trafo}t=\tt-h(\tr),\ee where $h(\tr)$ is called the
height function. In this new time coordinate, the quotient metric
$\tq$ can be written as
\[ \tq = \tq_{\tt\tt}\,dt^2+2\, \tq_{\tt\tt}\,h'\,dt\,d\tr +
(\tq_{\tt\tt}\,h'^2+\tq_{\tr\tr})\,d\tr^2, \quad \mathrm{where} \quad
h':=\frac{dh}{d\tr}. \] All metric components are functions of $\tr$
only. A major advantage of this manifest time-independence of the
metric functions is the implied simplification in numerical applications. 

The next question is how to choose the height function to obtain a
hyperboloidal foliation. As the time slices are required to be
spacelike and to approach future null infinity, the height function
should lead to a "bending up" of the surfaces as $\tr\to\infty$ such
that, asymptotically, their behavior is similar to outgoing null
surfaces. In the next subsection we will derive a necessary requirement
on the asymptotic behavior of the height function for the regularity
of the conformal compactification.

\subsection{Scri-fixing conformal compactification}
On level sets of the hyperboloidal time function $t$ we introduce a
compactifying coordinate $r$. The physical metric $\tg$ then becomes
singular at the domain boundary of the compactifying coordinate. To
compensate for this singular behavior we rescale the metric with a
conformal factor $\Omega$ that vanishes at that boundary in a suitable
way. This procedure is referred to as conformal compactification.

It is natural and convenient to require that the compactifying
coordinate $r$ is an area radius on the conformal extension with
respect to the rescaled metric $g=\Omega^2\,\tg$. This requirement
implies $\tr=r/\Omega$. The zero set of the conformal factor
corresponds to the asymptotic region of the physical spacetime.
Writing the rescaled quotient metric $q=\Omega^2\tq$ in terms of the
compactifying coordinate $r$ using $\tr(r)=r/\Omega(r)$, we get
\[ q = \Omega^2\,\tq_{\tt\tt}\,dt^2 + 2\, \tq_{\tt\tt}\,h'\,(\Omega -
\Omega'\,r)\, dt\,dr + \frac{\tq_{\tt\tt}\,h'^2 +
  \tq_{\tr\tr}}{\Omega^2}\,(\Omega-\Omega'\,r)^2 \,dr^2. \] This
formula can be applied with almost any choice of conformal
factor. Note that we have some freedom here. We can require, for
example, that $r$ be the proper distance in the conformal extension,
i.e.~$q_{rr}=1$. Then the conformal factor is determined by this
requirement and cannot, in general, be written in explicit form
\cite{Moncrief06}.  In contrast, by allowing the metric component
$q_{rr}$ to be more general, we are free to prescribe the
representation of a conformal factor in terms of a suitable
compactifying radial coordinate $r$ and the coordinate transformation
is explicit.

A convenient representation for the conformal factor that we will
resort to is $\Omega=(1-r)$. Our discussion is not restricted by this
choice. Other coordinate representations of the conformal factor are
also allowed. The above choice is not good at the origin as the
gradient of $r$ vanishes at this point. A suitable representation can
be used in the interior as will be discussed in the next
subsection. The above choice fixes the compactifying coordinate $r$
via \be \label{eq:comp_coord}\tr(r) = \frac{r}{1-r}, \ee implying
\be\label{eq:scrifixed} q = \Omega^2\,\tq_{\tt\tt}\,dt^2 + 2\,
\tq_{\tt\tt}\,h'\, dt\,dr + \frac{\tq_{\tt\tt}\,h'^2 +
  \tq_{\tr\tr}}{\Omega^2} \,dr^2. \ee The vanishing of $q_{tt}$ at
$\{\Omega=0\}$ implies that $\partial_t$ becomes null at this
surface. For regularity of the metric we require
\be\label{eq:height_asym} h' =
\sqrt{\frac{C\,\Omega^2-\tq_{\tr\tr}}{\tq_{\tt\tt}}} \qquad
\mathrm{as} \quad \tr\to\infty, \quad \mathrm{with} \quad
C=\mathrm{const.} \ee To gain an intuition for the behavior of
hyperboloidal surfaces, we will calculate coordinate speeds of
characteristics. The coordinate speed $c$ of ingoing and outgoing null
rays in radial directions can be calculated by
$c_{\mp}=(-q_{tr}\pm\sqrt{q_{tr}^2-q_{rr}q_{tt}})/q_{rr}$.  The speeds
are independent of the conformal rescaling of the metric which is a
reflection of the fact that conformal rescaling leaves the null cone
structure invariant.  The positive sign on the right hand side
corresponds to ingoing null rays, the negative sign corresponds to
outgoing null rays. On the conformal boundary we get $c_{\mp} =
(-q_{tr}\pm q_{tr})/q_{rr}$. As expected, the ingoing characteristic
speed is zero. Consequently no analytic boundary treatment is needed
or allowed. The outgoing characteristic speed is
$c_+|_{\scri^+}=-2q_{tr}/q_{rr}$. In a future hyperboloidal foliation
we have $q_{tr}|_{\scri^+}<0$ and $q_{rr}>0$, so that
$c_+|_{\scri^+}>0$ and radiation can leave the spacetime through
future null infinity.

\subsection{Matching}
An important feature of hyperboloidal surfaces, in contrast to
characteristic ones, is that their distinctive feature is determined
in the asymptotic region. While being spacelike or null is a local
property, being hyperboloidal or Cauchy is a global property. This
fact allows us to construct useful hyperboloidal foliations easily by
a matching procedure.

Foliations commonly used in the 3+1 approach are characterized by
properties which imply that the hypersurfaces approach spatial
infinity in the asymptotic domain. One may wish to keep these
properties in the interior and transform the surfaces in the exterior
such that they approach null infinity asymptotically.  Hyperboloidal
foliations can be constructed that coincide with arbitrary spacelike
foliations in the interior. To achieve this, we will employ the
following smooth transition between interior and exterior height
functions denoted by $h_i$ and $h_e$ respectively
\be \label{eq:transition} h(\tr) = \left\{ \begin{array}{ll} h_i(\tr)
  & \mathrm{for} \quad \tr\leq \tr_i,
  \\ e^{-(\tr-\tr_i)^2/(\tr-\tr_e)^2} (h_i(\tr)-h_e(\tr)) +
  h_e(\tr)\qquad & \mathrm{for} \quad \tr_i<\tr<\tr_e, \\ h_e(\tr) &
  \mathrm{for} \quad \tr\geq \tr_a. \end{array}\right.\ee The height
function needs to satisfy $|h'|<\sqrt{\tq_{\tr\tr}/(-\tq_{\tt\tt})}$
for the foliation to be spacelike. We may apply the above transition
also with respect to derivatives of the corresponding height
functions. We may set the conformal factor to unity in the interior
with a smooth transition to its asymptotic representation in the
exterior. The transition domain $(\tr_i,\tr_e)$ should be in the
exterior, but not necessarily in the far-field zone. The free
parameters of the matching can be decided upon by numerical
experiments. Detailed examples will be discussed in later sections.

One should note that this matching is fundamentally
different from Cauchy-characteristic matching where one matches a
spacelike and a null surface.  The matching in our case results in a
single, smooth, hyperboloidal surface with an arbitrary behavior in
the interior determined by $h_i(\tr)$.

\section{Minkowski spacetime}\label{sec:mink}
\subsection{Cauchy foliation}
\begin{figure}[ht]
   \centering 
   \begin{minipage}[t]{0.43\textwidth}
     \flushright 
    \psfrag{ip}{$i^+$} \psfrag{im}{$i^-$} \psfrag{i0}{$i^0$}
     \psfrag{scrp}{$\scri^+$}\psfrag{scrm}{$\scri^-$}
     \includegraphics[height=0.21\textheight]{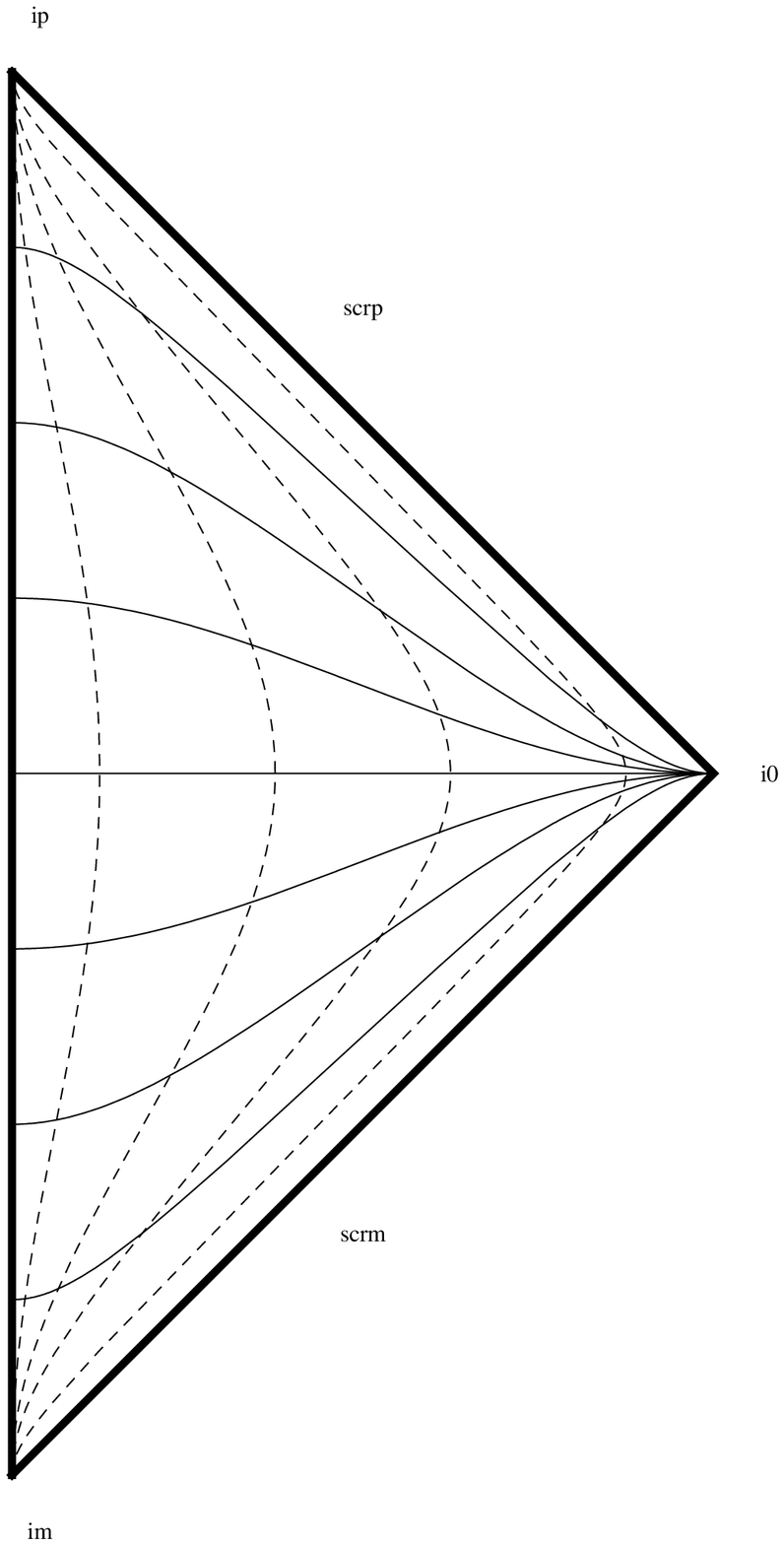}\hspace{1cm}
     \caption{Penrose diagram of Minkowski spacetime. Solid lines
       represent $\tt=\mathrm{const.}$ surfaces, dashed lines represent
       $\tr=\mathrm{const.}$ surfaces \label{fig:1}}
   \end{minipage}\hfill
   \begin{minipage}[t]{0.53\textwidth}
     \centering 
     \psfrag{t}{$\tt$} \psfrag{r}{$r$}
     \includegraphics[height=0.2\textheight]{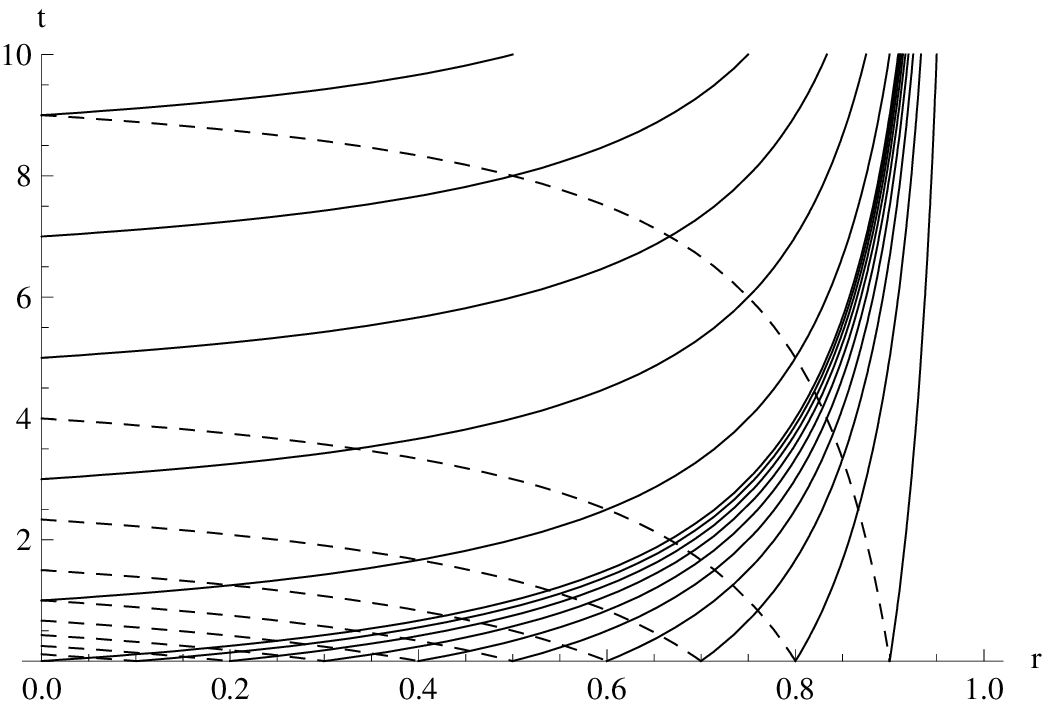}
     \caption{Characteristics in Minkowski spacetime in the standard
       foliation with respect to the compactifying coordinate $r$
       defined in (\ref{eq:comp_coord}). In- and outgoing null
       surfaces are represented by dashed and solid lines
       respectively. Note the different scales in $\tt$ and
       $r$. \label{fig:2}}
   \end{minipage}
 \end{figure}

The 2-dimensional Minkowski metric $\tilde{\eta}$ on the quotient space
in standard coordinates is given by
$\tilde{\eta}=-d\tilde{t}^2 + d\tilde{r}^2$.  The Penrose diagram in
Fig.~\ref{fig:1} depicts the causal structure of Minkowski
spacetime. Each point in the diagram represents a sphere except the
point $i^0$ that corresponds to spatial infinity and the
points along the vertical line segment connecting $i^+$ and
$i^-$ that correspond to the origin $\{\tr=0\}$.  Radial null
surfaces are represented by straight line segments with 45 degrees to
the horizontal. Curves with a smaller angle represent spacelike
surfaces, curves with a larger angle represent timelike
surfaces. Level surfaces of $\tilde{r}$ approach timelike infinity
denoted by $i^{\pm}$, level surfaces of $\tilde{t}$ approach spatial
infinity denoted by $i^0$.

It is evident from Fig.~\ref{fig:1} that an outgoing null ray never leaves
the surfaces of a Cauchy foliation. This is demonstrated more clearly
by the characteristics with respect to coordinates of the standard
foliation. We introduce on the Cauchy surfaces the compactifying
coordinate $r$ defined in (\ref{eq:comp_coord}). The resulting metric
reads $\tilde{\eta}=-d\tilde{t}^2 + dr^2/(1-r)^4$, and the coordinate
speeds of characteristics become \mbox{$c_{\pm} = \pm (1-r)^2$}.  The
characteristic speeds decrease to zero in the asymptotic region close
to $r=1$. In simulations, this results in a slowing down of outgoing
waves.  This feature is clearly shown in Fig.~\ref{fig:2} where in-
and outgoing characteristics are depicted by dashed and solid lines
respectively.  We see that outgoing characteristics pile up during the
evolution near the boundary and eventually, cannot be resolved (see
\cite{Pretorius05a} for a discussion of this phenomena in the context
of a binary black hole evolution).

A suitable compactified foliation should not only have no ingoing
characteristics coming in from the outer boundary, but also allow
outgoing characteristics to leave the simulation domain.
\subsection{Hyperboloidal foliations}
\begin{figure}[ht]
  \centering
  \psfrag{ip}{$i^+$}
  \psfrag{im}{$i^-$}
  \psfrag{i0}{$i^0$}
  \psfrag{scrp}{$\scri^+$}
  \psfrag{scrm}{$\scri^-$}
  \includegraphics[height=0.21\textheight]{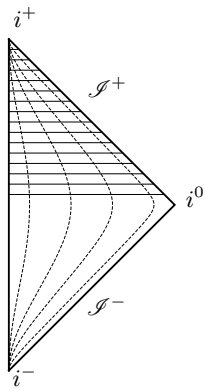}
  \caption{Penrose diagram of a hyperboloidal foliation of
    Minkowski spacetime. Solid lines represent hyperboloidal
    surfaces, dashed lines represent Killing observers \label{fig:3}}
\end{figure}

A typical example for a future hyperboloidal foliation is plotted in
Penrose diagram Fig.~\ref{fig:3}. Null rays do leave the surfaces of a
hyperboloidal foliation. The corresponding time function to the
foliation in Fig.~\ref{fig:3} is given for $\tt > 0$ by
\be \label{eq:typhyp}t(\tt,\tr)=\frac{1}{2
  \tt}\left(\tt^2-\tr^2-1\right). \ee A distinctive feature of
hyperboloidal surfaces is that their mean extrinsic curvature attains
a non-vanishing, finite value at infinity. We use the convention in
which positive $\tilde{K}$ corresponds to expansion so that positive
asymptotic values of $\tilde{K}$ correspond to future hyperboloidal
surfaces. The mean extrinsic curvature of level sets of $t$ as defined
in (\ref{eq:typhyp}) reads $\tilde{K}=3/\sqrt{1+t^2}$. Each surface
has a spatially constant mean curvature.  The foliation, however, does
not seem to be convenient for numerical calculations, because the time
function does not have the form (\ref{eq:trafo}). The Minkowski metric
in this new coordinate depends on time.

\begin{figure}[ht]
  \centering
  \psfrag{ip}{$i^+$}
  \psfrag{im}{$i^-$}
  \psfrag{i0}{$i^0$}
  \psfrag{scrp}{$\scri^+$}
  \psfrag{scrm}{$\scri^-$}
  \includegraphics[height=0.21\textheight]{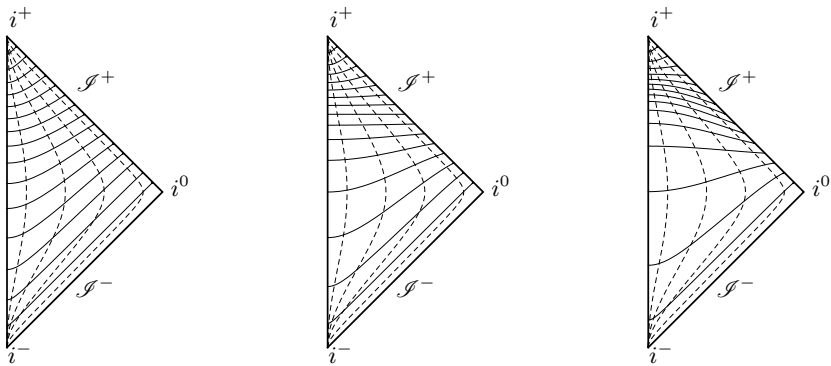}
  \caption{Penrose diagrams for constant mean curvature (CMC)
    foliations of Minkowski spacetime. Solid lines represent
    hyperboloidal surfaces, dashed lines represent Killing
    observers. The foliations have $\tilde{K}=\{6,3,2\}$ from left to
    right. \label{fig:4}}
\end{figure}

A more convenient foliation can be constructed by Lie dragging a
standard hyperboloid along the timelike Killing field.  The
spherically symmetric standard hyperboloids in Minkowski spacetime are
given by $a^2=\tt^2-\tr^2,$ with $a\in\mathbb{R}_+$. Translating these
surfaces along $\partial_{\tilde{t}}$ by an amount of $t$, we get the
equation $a^2=(\tt-t)^2-\tr^2$. The new time coordinate $t$ is then
defined by \be\label{eq:sthyp} t(\tt,\tr)=\tt-\sqrt{a^2+\tr^2}. \ee
Each surface of the foliation has a constant mean curvature $\tilde{K}
= 3/a$. The Penrose diagram in Fig.~\ref{fig:4} shows foliations
defined by (\ref{eq:sthyp}) for different values of $\tilde{K}$.

In this new time coordinate, scri-fixing conformal compactification as in
(\ref{eq:scrifixed}) results in 
\be \label{eq:cmc_mink}\eta=-\Omega^2dt^2-
\frac{2r}{\sqrt{a^2\,\Omega^2+r^2}}\,dt dr +
\frac{a^2}{a^2\,\Omega^2+r^2}\,dr^2. \ee 

\begin{figure}[ht]
  \centering
  \includegraphics[height=0.15\textheight]{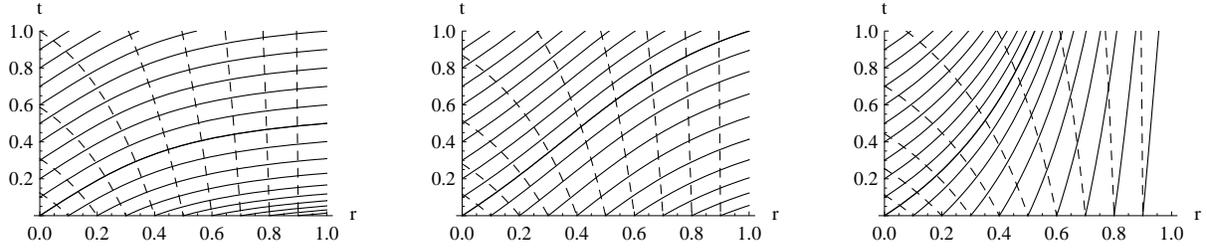}
  \caption{Coordinate representations of in- and outgoing
    characteristics of (\ref{eq:cmc_mink}) with
    $\tilde{K}=3/a=\{6,3,0.5\}$. High values for $\tilde{K}$ are seen
    to be restrictive in a numerical calculation in which the
    numerical cone needs to be larger than the characteristic
    cone. \label{fig:5}}
\end{figure}

We see from the Penrose diagram Fig.~\ref{fig:4} that the mean
extrinsic curvature of the surfaces determines the angle at which the
surfaces cut null infinity. It also acts as a measure for how close
the surfaces are to outgoing null surfaces. The similarity of future
hyperboloidal surfaces to outgoing null surfaces can be quantified by
the coordinate speed of the outgoing characteristic at null infinity.
We have $c_+|_{\scri^+}=4 \tilde{K}^2/9$. A large value of $\tilde{K}$
corresponds to surfaces which are closer to null surfaces, as also
suggested by the Penrose diagrams in Fig.~\ref{fig:4}. This is seen
more clearly in Fig.~\ref{fig:5}, where in- and outgoing
characteristics have been plotted with respect to coordinates. Note
that the choice of a high value for $\tilde{K}$ in a numerical
calculation using an explicit time integration algorithm would require
very small time steps due to the Courant condition which demands that
the numerical cone needs to be larger than the characteristic
cone. For small values of $\tilde{K}$, however, this condition can be
easily satisfied as seen also in Fig.~\ref{fig:5}.

It should be emphasized that the notion of "proximity" to null
surfaces does not have a general geometric meaning. Geometrically, it
makes no sense to claim that a spacelike surface is close to a null
surface as this depends on the chosen frame of reference. On the other
hand, in stationary spacetimes, we have a natural choice of observers,
namely the Killing observers, so that we can introduce such a
notion. In the general non-stationary case, this notion may be
useful in the asymptotic domain with respect to the asymptotic Killing
observers.
\subsection{Matching}
\begin{figure}[ht]
  \centering \psfrag{ip}{$i^+$} \psfrag{im}{$i^-$} \psfrag{i0}{$i^0$}
  \psfrag{scrp}{$\scri^+$}\psfrag{scrm}{$\scri^-$}
  \begin{minipage}[t]{0.47\textwidth}
    \flushright 
    \includegraphics[height=0.21\textheight]{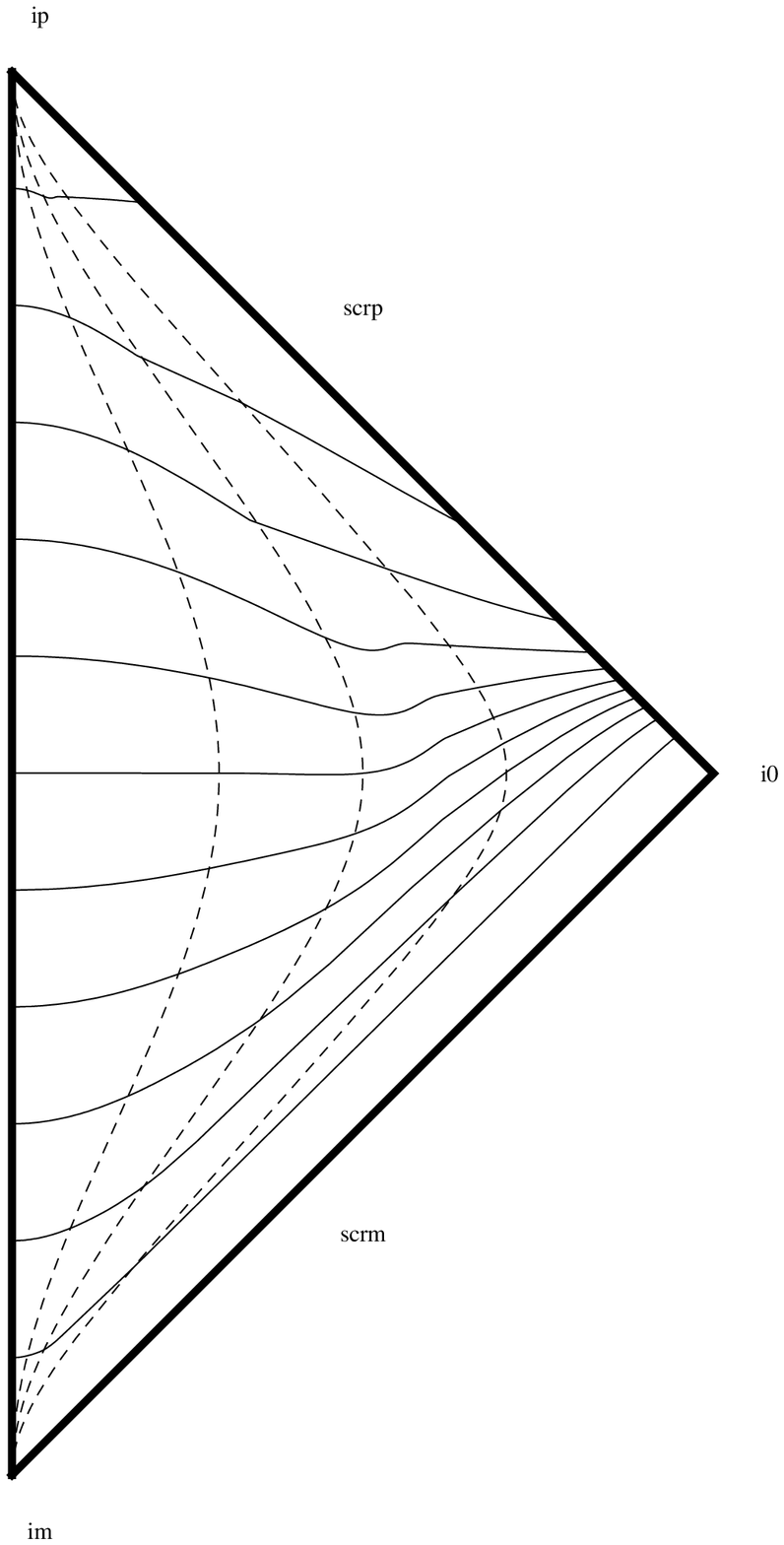}\hspace{1cm}
  \end{minipage}\hfill
  \begin{minipage}[t]{0.47\textwidth}
    \centering
    \includegraphics[height=0.21\textheight]{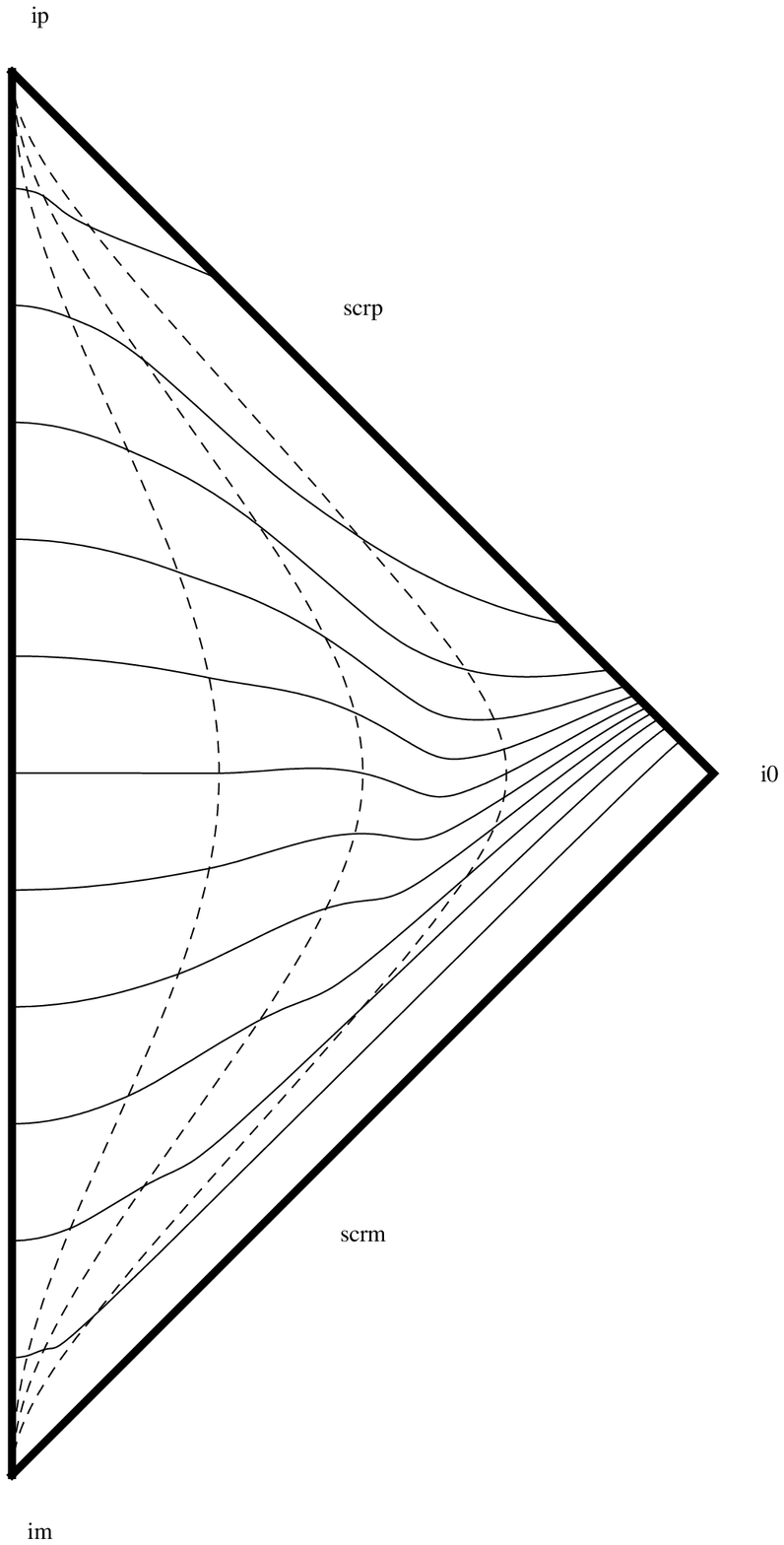}
  \end{minipage}
  \caption{Transition between maximal surfaces and two types of
    hyperboloidal surfaces determined by (\ref{eq:transit1}) and
    (\ref{eq:transit2}). The three dashed lines depict the timelike surfaces
    $\tr_i=1/2,\ \tr_m=1$, and $\tr_e=2$. The exterior foliation has
    $a=2$ in both cases. \label{fig:6}}
\end{figure}

Level surfaces of the standard time coordinate $\tt$ in Minkowski
spacetime are maximal, i.e.~their mean extrinsic curvature
vanishes. To match an interior maximal foliation to a hyperboloidal
foliation, we choose in (\ref{eq:transition}) \be\label{eq:transit1}
h_i = C_i, \qquad h_e = C_e + \sqrt{a^2+\tr^2}.\ee The constants $C_i$
and $C_e$ must be chosen such that the matched surface is spacelike in
the transition domain, that is, $|h'|<1$. One convenient way to achieve
this is to require that the surfaces intersect within the transition
domain. We denote this matching point by $\tr_m\in (\tr_i,\tr_e)$, and
require $h_i(\tr_m)=h_e(\tr_m)$. This implies
\mbox{$C_e=C_i-\sqrt{a^2+\tr_m^2}$}. The resulting foliation has been
plotted in Fig.~\ref{fig:6}. The parameters have been chosen such that
the matching is clearly visible in the diagrams. In a numerical
calculation, they need to be decided upon by numerical experience.

A more general way of constructing hyperboloidal surfaces in a given
spacetime is to Taylor expand the formula (\ref{eq:height_asym}) near
infinity. We get in Minkowski spacetime \be\label{eq:transit2}
h'(\tr)\sim \left(1-\frac{a^2}{(1+\tr)^2}\right)^{\frac{1}{2}} =
1-\frac{a^2}{2 \tr^2} + \frac{a^2}{\tr^3} +
O\left(\frac{1}{\tr^4}\right), \quad \mathrm{as} \ \tr\to\infty,\ee
where we set $C\equiv a^2$. The first two terms of the expansion are
sufficient to fulfill the requirements on the height function for a
future hyperboloidal foliation. The foliation is defined in the
exterior domain for $\tr>a-1$, or written in the compactifying
coordinate, for $r>1-1/a$. We use (\ref{eq:transition}) to match the
surfaces to an interior maximal foliation by choosing $h_e(\tr)=\tr +
a^2/(2 \tr)+C_e$, and $h_i=C_i$. The resulting foliation has been
plotted in Fig.~\ref{fig:6}. The rescaled metric in the exterior
domain reads
\[ \eta =  -\Omega^2\,dt^2-2\left(1-\frac{a^2\Omega^2}{2 r^2}\right)
\,dt dr + \frac{a^2}{2r^2}\,dr^2. \]
The outgoing characteristic speed at future null infinity is
$c_+|_{\scri^+}=4/a^2$. The mean extrinsic curvature of these surfaces
depends on $r$ and attains the value $\tilde{K}=3/a$ at
infinity. 

\section{Schwarzschild spacetime}\label{sec:ss}
\subsection{Cauchy-type foliations}
All foliations of Schwarzschild spacetime commonly employed in
numerical calculations within the 3+1 approach are Cauchy-type
foliations approaching spatial infinity in the asymptotic domain. We
mention some of them in this subsection for later reference. We focus
on the physically interesting part of the Schwarzschild-Kruskal
extension (the regions I and II in \cite{Hawking73}) which we refer to
as the extended Schwarzschild spacetime.
\subsubsection{Schwarzschild coordinates}
The 2-dimensional Schwarzschild metric in Schwarzschild coordinates is
given by \be\label{eq:st_ss} \tilde{q} = -\left(1-\frac{2
  m}{\tr}\right)\,d\tt^2 + \left(1-\frac{2m}{\tr}\right)^{-1}d\tr^2,
\qquad \tr>2m.\ee Level surfaces of $\tt$ are maximal. They intersect at the
bifurcation point and at spatial infinity as seen in
Fig.~\ref{fig:7}. Therefore, these coordinates are suitable neither
to study the region near the black hole, nor the asymptotic domain.
\subsubsection{Eddington-Finkelstein coordinates}
\begin{figure}[ht]
  \centering
  \psfrag{sing}{\footnotesize{singularity}} \psfrag{hor}{$\mathcal{H}$}
  \psfrag{ip}{$i^+$} \psfrag{im}{$i^-$} \psfrag{i0}{$i^0$}
  \psfrag{scrp}{$\scri^+$} \psfrag{scrm}{$\scri^-$}
  \begin{minipage}[t]{0.31\textwidth}
  \includegraphics[width=\textwidth]{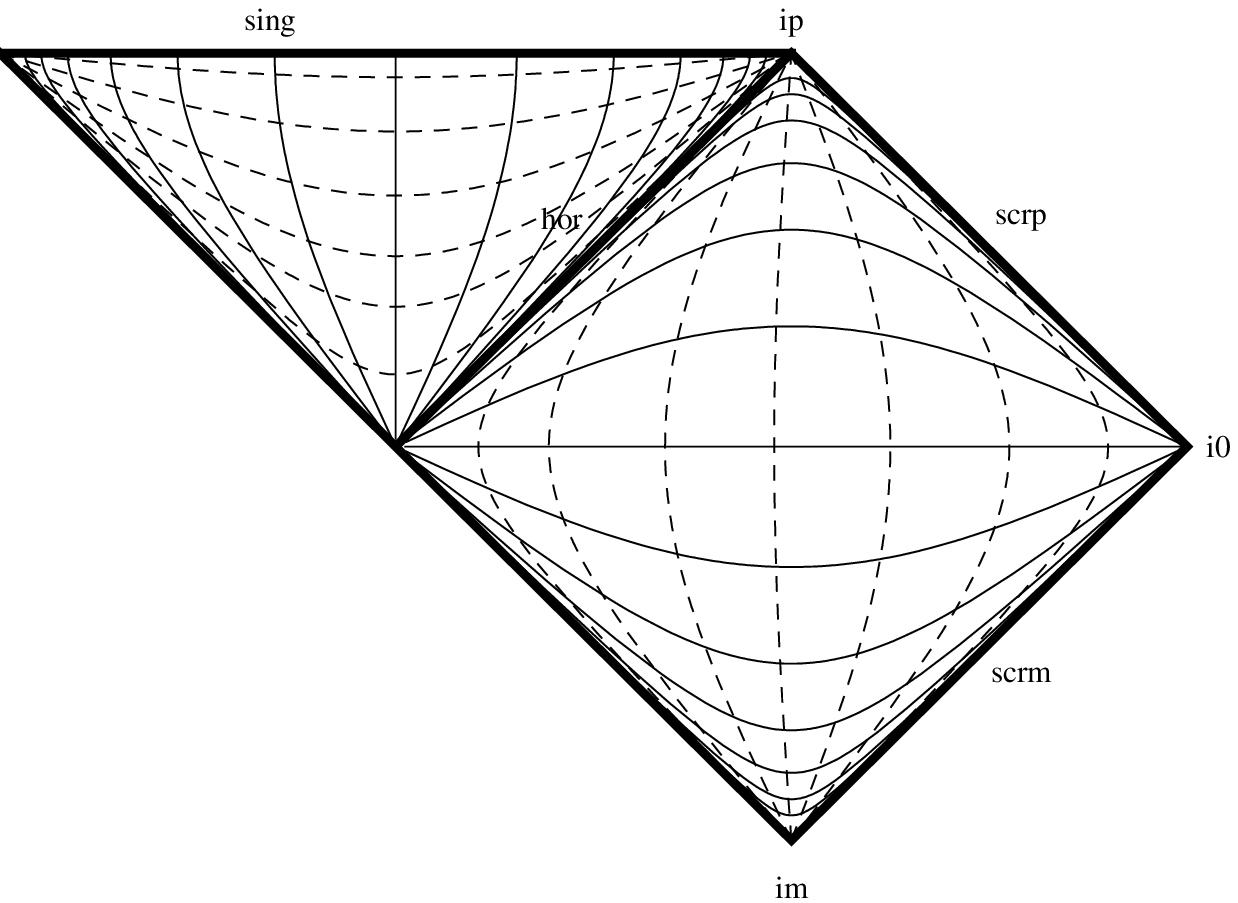}
  \caption{Level sets of Schwarzschild coordinates on the extended
    Schwarzschild manifold.\label{fig:7}}
   \end{minipage}\hfill
   \begin{minipage}[t]{0.31\textwidth}
     \centering 
     \includegraphics[width=\textwidth]{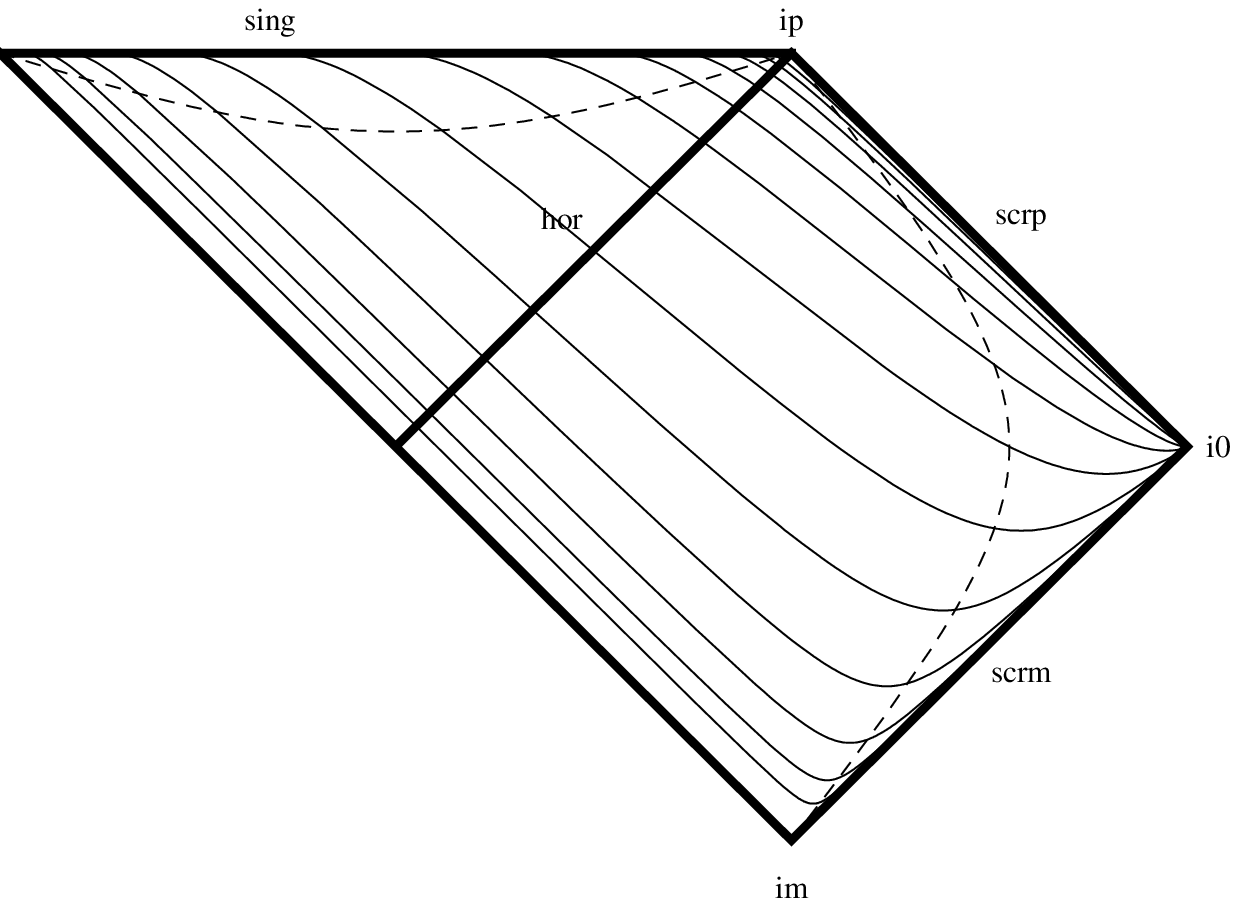}
     \caption{Eddington-Finkelstein foliation suitable for the
       excision technique. Dashed lines correspond to \mbox{$\tr=\{1.5 m,
       4m\}$}. \label{fig:8}}
   \end{minipage}\hfill
   \begin{minipage}[t]{0.31\textwidth}
     \centering 
     \includegraphics[width=\textwidth]{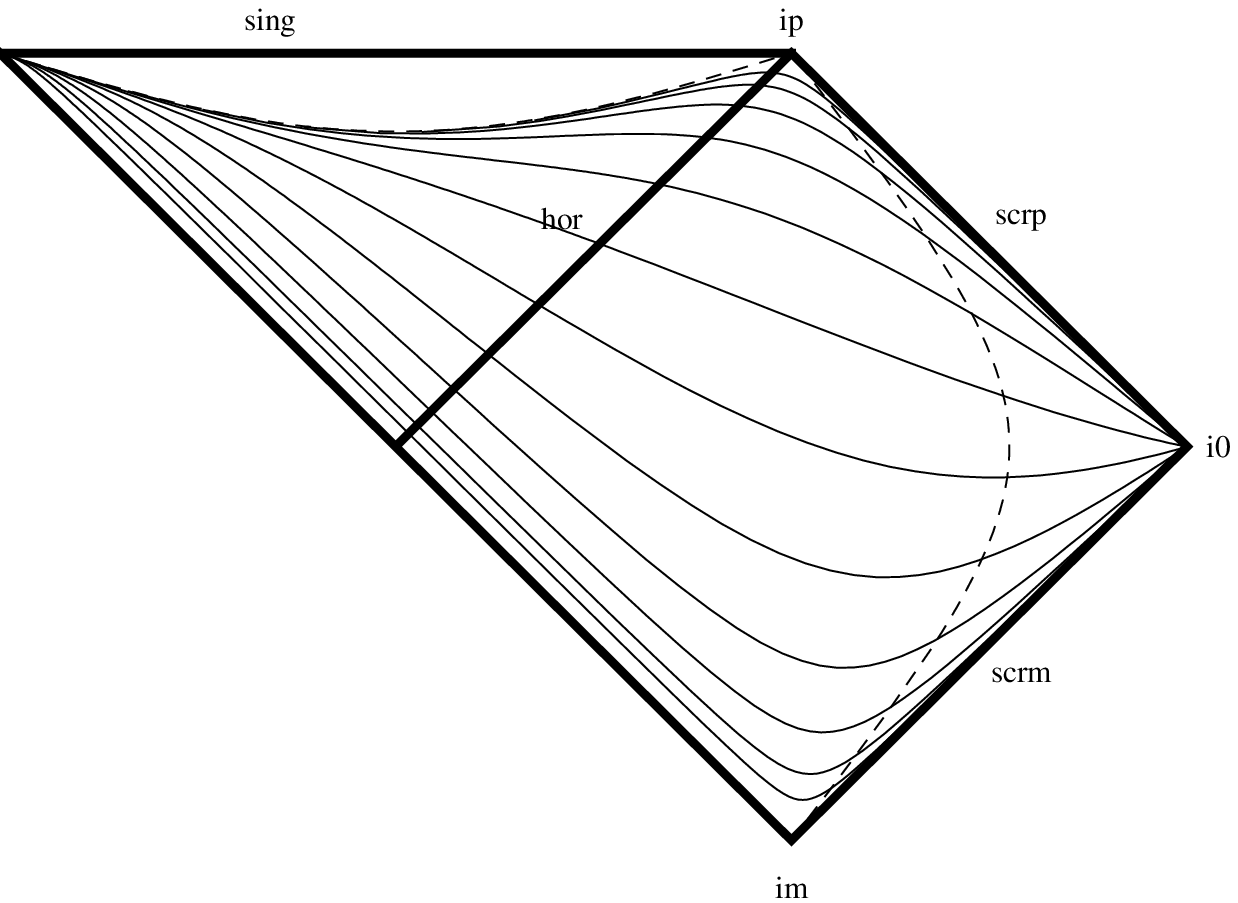}
     \caption{Maximal slices suitable for the moving puncture
       technique. Dashed lines correspond to \mbox{$\tr=\{1.5 m,
         4m\}$}. \label{fig:9}}
   \end{minipage}
\end{figure}

In numerical applications, a common choice of so-called "horizon penetrating"
coordinates in Schwarzschild spacetime is defined as in
(\ref{eq:trafo}) with the height function
\[h(\tr)= 2m\,\ln\,\left|\frac{\tr}{2m}-1\right|. \]
The coordinates such defined have originally been discussed by
Eddington \cite{Eddington24} and Finkelstein \cite{Finkelstein58}. The
Schwarzschild metric on the extended Schwarzschild manifold takes the
form \be \label{app:ief} \tq=-\left(1-\frac{2m}{\tr}\right)\,dt^2 +
\frac{4m}{\tr} \, dt d\tr+\left( 1 + \frac{2m} {\tr}\right) \, d\tr^2,
\qquad \tr>0. \ee The mean extrinsic curvature of level sets of $t$
depends on $\tr$ and vanishes in the asymptotic domain.  The foliation
by the time function $t$ is depicted in the Penrose diagram
Fig.~\ref{fig:8}.  We see that the foliation penetrates the horizon
which makes it suitable for excising the singularity from the
computational domain. The surfaces are spacelike at each point. They
approach spatial infinity in the asymptotic domain which makes them
inconvenient to study radiation.

\subsubsection{Maximal slices for the puncture technique}
Currently, the most popular approach in numerical calculations of
black hole spacetimes is the moving puncture technique. The evolution
with puncture initial data for a Schwarzschild black hole in a
modified 1+log slicing reaches a final stationary state that
corresponds to a maximal slice given in suitable coordinates
\cite{Hannam06a, Hannam06b, Baumgarte07, Hannam:2008sg}. We will use
maximal foliations to demonstrate the possibility that hyperboloidal
surfaces can be constructed which are suitable for the application of
the moving puncture technique in the interior.

The requirement that the mean extrinsic curvature $\tilde{K}$ of the
surfaces defined as in (\ref{eq:trafo}) are constant implies in
Schwarzschild spacetime \be\label{eq:cmcK} \tilde{K} =
\frac{1}{\tilde{r}^2} \partial_{\tilde{r}}\left(\frac{\tilde{r}^2
  h'\left(1-\frac{2m} {\tilde{r}}\right)}
     {\sqrt{\left(1-\frac{2m}{\tilde{r}} \right)^{-1} -
         h'^2\left(1-\frac{2m}{\tilde{r}}\right)}}\right).\ee We can
     integrate this relation once for a maximal foliation
     ($\tilde{K}=0$) in Schwarzschild spacetime to obtain
     \cite{Estabrook73, Beig97} \be\label{eq:maximal} h'(\tr) = -
     \frac{C}{\left(1-\frac{2m}{\tr}\right) \sqrt{\tr^4-2 m \tr^3 +
         C^2}}, \ee where $C$ is a constant of integration. The
     Schwarzschild metric becomes
\[\tq=-\left(1-\frac{2m}{\tr}\right)\,dt^2 + 
\frac{2 C}{f \tr^2} \, dt d\tr+ \frac{1}{f^2} \, d\tr^2, \qquad
\mathrm{with} \qquad f = \sqrt{1-\frac{2m}{\tr}+\frac{C^2}{\tr^4}}.\]
The standard foliation of Schwarzschild spacetime is a member of this
family with $C=0$. The final stationary state for a puncture evolution
in Schwarzschild spacetime is similar to a limiting maximal slice with
$C=3\sqrt{3}\,m^2/4$. Fig.~\ref{fig:9} shows a maximal foliation with
$C=3\sqrt{3}\,m^2/4$ in the Penrose diagram of the extended
Schwarzschild spacetime together with the limiting surface at
$\tr=3m/2$ where $f=0$.
\subsection{Hyperboloidal CMC-foliations}
The main difference between the Schwarzschild foliation in
Fig.~\ref{fig:7} and the horizon penetrating foliations in
Fig.~\ref{fig:8} and Fig.~\ref{fig:9} is the behavior of surfaces near
the event horizon. The intersection at the bifurcation point in
Schwarzschild coordinates is not suitable to study the black
hole. Similarly, the intersection of surfaces of a foliation at
spatial infinity is not suitable to study radiation near null
infinity. The construction of hyperboloidal foliations, while
geometrically quite different, can in this respect be regarded as
similar to the construction of horizon penetrating coordinates.

\begin{figure}[ht]
  \centering 
  \begin{minipage}[t]{0.47\textwidth}
    \flushright \psfrag{sing}{\footnotesize{singularity}}
    \psfrag{hor}{$\mathcal{H}$} \psfrag{ip}{$i^+$} \psfrag{im}{$i^-$}
    \psfrag{i0}{$i^0$} \psfrag{scrp}{$\scri^+$}
    \psfrag{scrm}{$\scri^-$}
    \includegraphics[height=0.21\textheight]{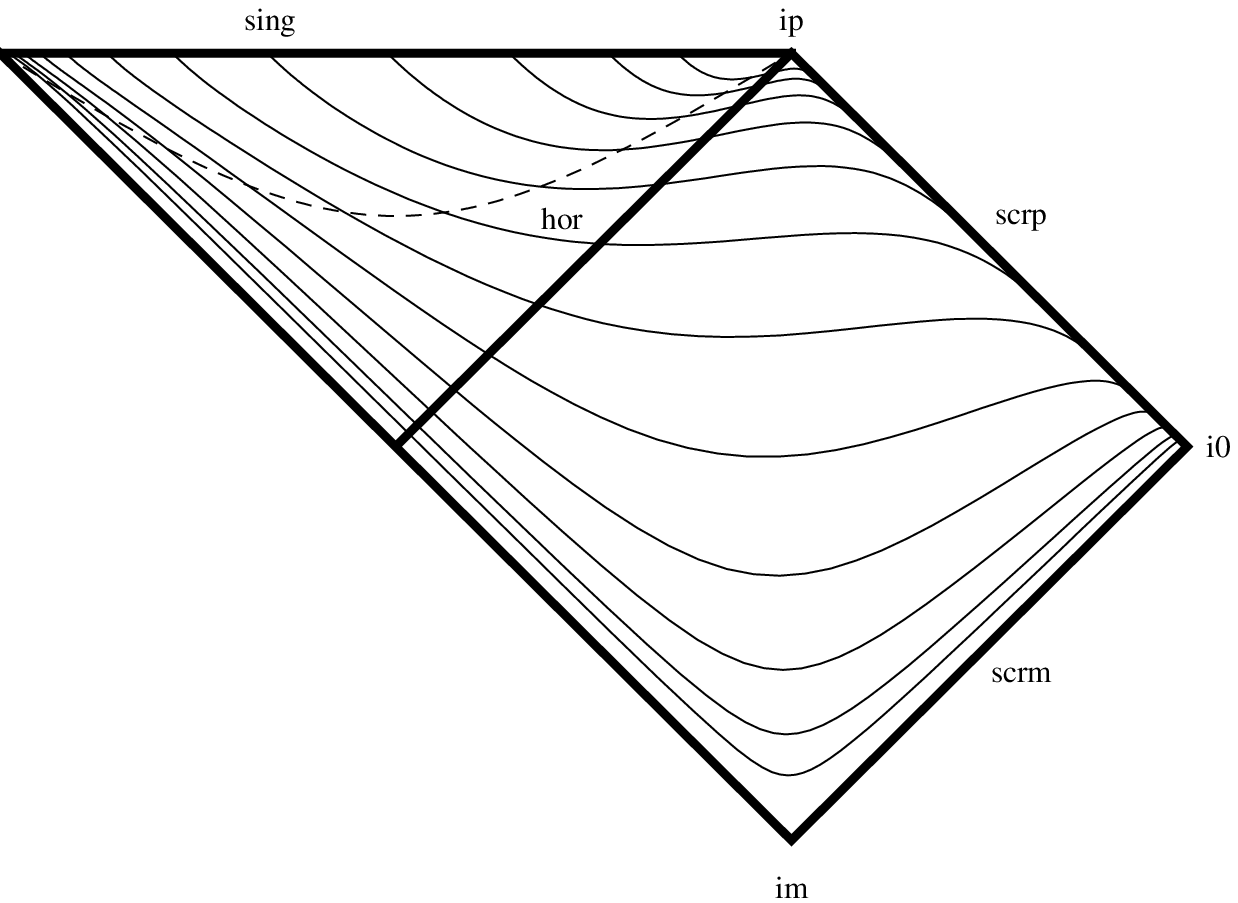}\hspace{1cm}
  \end{minipage}\hfill
  \begin{minipage}[t]{0.47\textwidth}
    \centering \psfrag{t}{$t$} \psfrag{h}{$\mathcal{H}$}
    \psfrag{r}{$r$} \psfrag{scr+}{$\scri^+$}
    \includegraphics[height=0.21\textheight,width=0.8\textwidth]{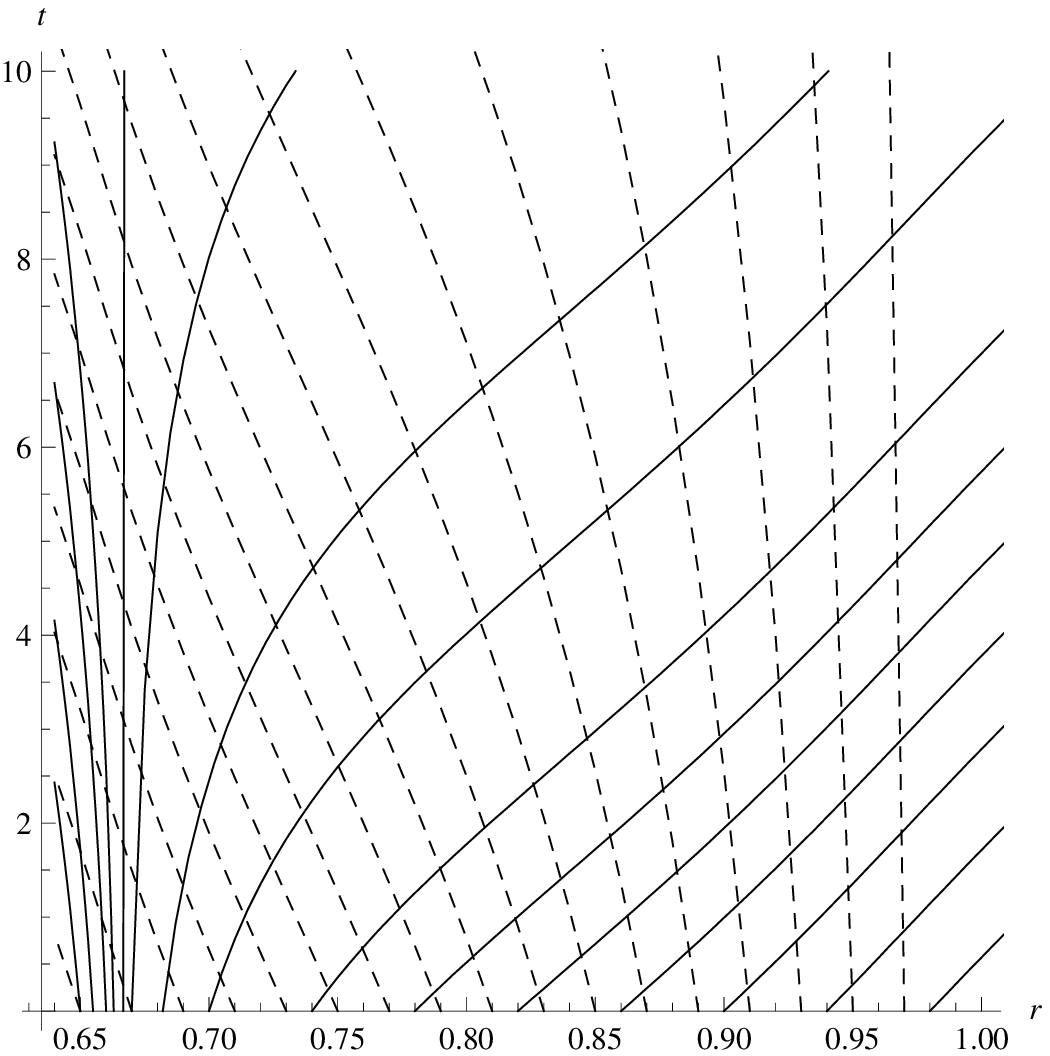}
  \end{minipage}
\caption{CMC-foliation of the extended Schwarzschild spacetime with
  $\tilde{K}=0.4$, $C=2$ and $m=1$. On the left, the Penrose diagram is
  plotted, on the right in- and outgoing characteristics. The
  excision surface is taken to be at $\tr=1.8m$ and is plotted as
  the dashed line in the Penrose diagram. The event horizon is at
  $r=2/3$. \label{fig:10}}
\end{figure}

To find a representation of Schwarzschild spacetime based on
hyperboloidal surfaces, we use the general family of spherically
symmetric CMC surfaces in the extended Schwarzschild spacetime as
discussed in \cite{Brill80, MalecMurch03} (see \cite{Gentle00} for a
numerical study).  The integration of (\ref{eq:cmcK}) with
$\tilde{K}\ne 0$ and integration constant $C$ leads to the
differential equation \be\label{app:height}
h'=\frac{\frac{\tilde{K}\tilde{r}^3}{3} - C}
{\left(1-\frac{2m}{\tilde{r}}\right)\tilde{P}(\tilde{r})}, \quad
\mathrm{with} \quad \widetilde{P}(\tr) :=
\sqrt{\left(\frac{\tilde{K}\tr^3}{3}-C\right)^2+
  \left(1-\frac{2m}{\tr}\right)\tr^4}.\ee The maximal foliation given
in (\ref{eq:maximal}) is a member of this family with $\tilde{K}=0$.
To get the height function we need to integrate the differential
equation in (\ref{app:height}). The integral cannot be written in
explicit form but needs to be carried out numerically. The resulting
foliation has been plotted in Fig.~\ref{fig:10}.

Scri-fixing conformal compactification as in (\ref{eq:scrifixed})
based on this foliation results in \be \label{cmc_ss} q =
-\left(1-\frac{2 m \,\Omega}{r}\right)\Omega^2 dt^2 -
\frac{2\left(\tilde{K}r^3/3-C\,\Omega^3\right)}{P(r)}\,dt dr +
\frac{r^4}{P^2(r)}\,dr^2, \ee where
\[ P(r):=\Omega^3\,\widetilde{P}(r) = \sqrt{\left(\frac{\tilde{K}r^3}{3} -
C\Omega^3\right)^2 + \left(1-\frac{2 m (1-r)}{r}\right)\Omega^2
r^4}.\] 

The global behavior of CMC-surfaces in Schwarzschild
spacetime depends on the free parameters as discussed in 
\cite{MalecMurch03}. The Penrose diagram in Fig.~\ref{fig:10} depicts
surfaces which are convenient for numerical calculations. They come
from future null infinity, pass the horizon above the bifurcation
sphere and run into the future singularity.  Depending on the free
parameters, $P(r)$ may have a non-vanishing zero set. The corresponding
surface lies always inside the event horizon and does not pose a
difficulty if the excision surface has been chosen carefully.

The outgoing characteristic speed at $\scri^+$ reads
$c_+|_{\scri^+} = 4 \tilde{K}^2/9$, just as for the CMC-foliation of
Minkowski spacetime. The main difference to Minkowski spacetime is the black
hole region. The coordinate representation of characteristics in
Fig.~\ref{fig:10} illustrates that there are no incoming
characteristics into the simulation domain both at the excision
boundary inside the event horizon and at the outer boundary on future
null infinity.

\subsection{Matching}
\begin{figure}[ht]
  \centering \psfrag{sing}{\footnotesize{singularity}}
  \psfrag{hor}{$\mathcal{H}$} \psfrag{ip}{$i^+$} \psfrag{im}{$i^-$}
  \psfrag{i0}{$i^0$} \psfrag{scrp}{$\scri^+$} \psfrag{scrm}{$\scri^-$}
  \begin{minipage}[t]{0.47\textwidth}
    \flushright
    \includegraphics[height=0.21\textheight]{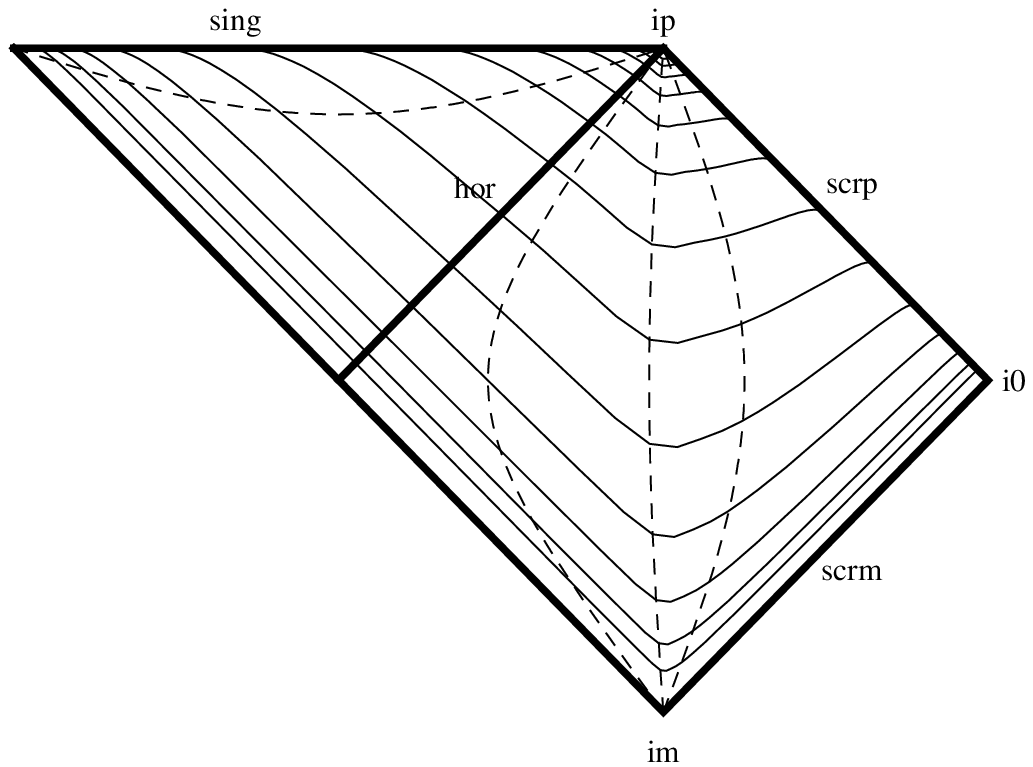}\hspace{1cm}
  \end{minipage}\hfill
  \begin{minipage}[t]{0.47\textwidth}
    \centering 
    \includegraphics[height=0.21\textheight]{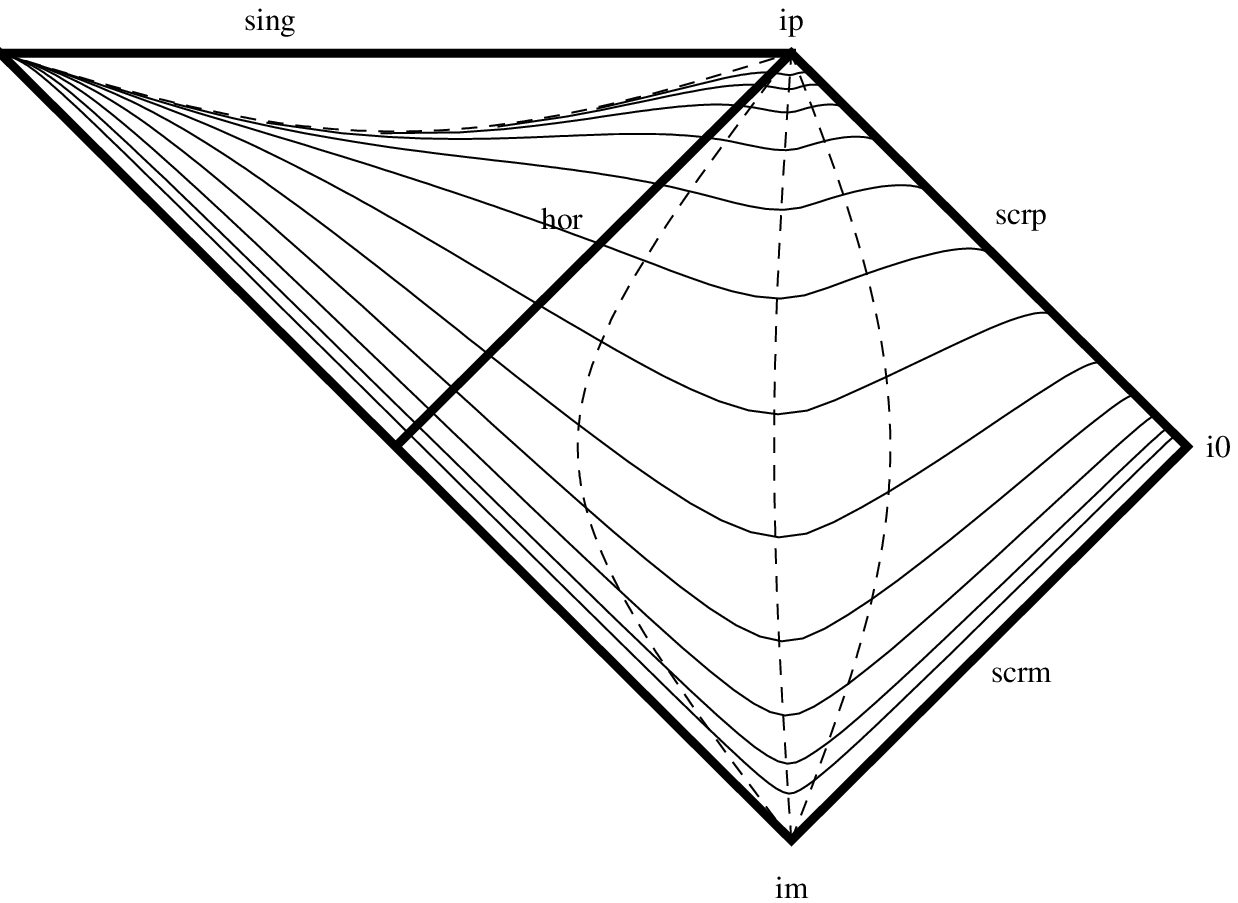}
  \end{minipage}
\caption{Penrose diagrams of matched foliations in the extended
  Schwarzschild spacetime. The dashed lines correspond to
  $\tr=1.5\,m,\ \tr_i=2.1\,m,\ \tr_m=2.5\,m,\ \tr_e=3\,m$. The diagram
  on the left illustrates matching between interior
  Eddington-Finkelstein surfaces and surfaces with height function
  $h=\sqrt{9+\tr_\ast^2}$. The diagram on the right illustrates
  matching between maximal slices and hyperboloidal surfaces with the
  height function (\ref{eq:asymp_hyp}). \label{fig:11}}
\end{figure}

Schwarzschild spacetime is asymptotically flat in null directions,
implying that its asymptotic structure is similar to Minkowski
spacetime. We can therefore use the same form of the height function
as we did for constructing a hyperboloidal foliation in Minkowski
spacetime to get, at least asymptotically, a hyperboloidal foliation
in Schwarzschild spacetime. In writing the height function, however,
we must be careful about the choice of coordinates. Asymptotic
flatness in null directions can be made manifest in terms of the
Regge-Wheeler coordinate $\tr_\ast=r+2m \ln(r-2m)$ as described in
\cite{Penrose80}. It is therefore the height function
$h(\tr)=\sqrt{a^2+\tr_\ast^2}$, that gives a hyperboloidal foliation
in the exterior domain. As the coordinate $\tr_\ast$ is not suitable
near the black hole region, we match the surfaces to interior
Eddington-Finkelstein surfaces. The resulting foliation has been
plotted in Fig.~\ref{fig:11}.

The simplest method to construct hyperboloidal surfaces seems to be
the Taylor expansion of (\ref{eq:height_asym}) as in
(\ref{eq:transit2}). In Schwarzschild spacetime, the expansion
delivers \be \label{eq:ss_taylor} h'(\tr)\sim
\frac{\tr}{\tr-2m}\, \left(1-\left(1-\frac{2m}{\tr}\right)
\frac{C}{(1+\tr)^2}\right)^{\frac{1}{2}} = 1+\frac{2m}{\tr} +
\frac{8m^2-C}{2 \tr^2} + \frac{C+8m^3}{\tr^3} +
O\left(\frac{1}{\tr^4}\right), \quad \mathrm{as}\ \tr\to\infty.\ee We
take the first two terms so that the height function becomes
\be\label{eq:asymp_hyp} h(\tr) = \tr+2m \ln \tr, \ee which essentially
corresponds to the Regge-Wheeler coordinate in the exterior
domain. Matching to an interior maximal foliation gives the second
conformal diagram in Fig.~\ref{fig:11}. The conformally rescaled
metric in a scri-fixing gauge in the exterior domain reads
\[ q = -\left(1-\frac{2m\Omega}{r}\right) \Omega^2\,dt^2 - 2
\left(1-\frac{4m^2\Omega^2}{r^2}\right)\, dt dr + \frac{8 m^2
  (r^2-2m^2\Omega^2)}{r^3(r - 2m\Omega)}\, dr^2.\] The outgoing
characteristic speed at future null infinity is $c_+|_{\scri^+} =
1/(2m^2)$. The mean extrinsic curvature of the time slices attains a
positive finite value at infinity that depends on the mass of the
Schwarzschild spacetime. It reads
$\tilde{K}|_{\scri^+}=3/(2\sqrt{2}m)$. To gain some control on
properties of this asymptotic foliation one can take further terms in
the Taylor expansion (\ref{eq:ss_taylor}).
\section{Kerr spacetime}\label{sec:kerr}
In most studies, going from the spherically symmetric case to the
axially symmetric case is a challenging task. Take as an example the
construction of suitable null coordinates. Null coordinates can easily
be constructed on Minkowski and Schwarzschild spacetimes. In Kerr
spacetime, however, they are difficult to deal with and cannot be
written in explicit form \cite{Bishop05,Pretorius98}. This is, in
part, a consequence of the fact that null coordinates are very rigid.
Spacelike foliations on the other hand are simpler to construct due to
the large freedom involved in their choices. Further, the property of
a spacelike surface to be hyperboloidal does not restrict its form in
the interior as we have seen in previous sections. It only determines
its asymptotic behavior. This fact in combination with the asymptotic
flatness of Kerr spacetime allows us to carry over the construction of
a scri-fixing gauge in spherical symmetry directly to axial symmetry
on the example of Kerr spacetime.

The most commonly used coordinates for Kerr spacetime in numerical
relativity are the Boyer-Lindquist and the Kerr-Schild coordinates. In
the following we present the asymptotic form of simple hyperboloidal
foliations in these coordinates. Other choices are possible and
straightforward using the techniques presented so far. The Kerr
metrics as given in (\ref{eq:kerr_bl}) and (\ref{eq:kerr_ks}) can be
used in the exterior domain. One may employ the matching technique to
obtain a convenient foliation also in the interior depending on the
problem at hand.
\subsection{In Boyer-Lindquist coordinates}
The Kerr metric in Boyer-Lindquist coordinates reads
\[ \tilde{g} = -\left(1-\frac{2m\tr}{\widetilde{\Sigma}}\right) d\tt^2 -  
\frac{4 a m \tr}{\widetilde{\Sigma}}\sin^2\!\vartheta \,d\tt\,d\varphi
+ \frac{\widetilde{\Sigma}}{\widetilde{\triangle}}\, d\tr^2 +
\widetilde{\Sigma}\,d\vartheta^2+\tilde{R}^2\sin^2\vartheta\,d\varphi^2,\]
where \be\label{eq:kerrdef} \widetilde{\Sigma} = \tr^2 + a^2
\cos^2\vartheta, \qquad \widetilde{\triangle} = \tr^2 + a^2 - 2m\tr,
\qquad \tilde{R}^2=\tr^2+a^2+
\frac{2ma^2\tr\sin^2\vartheta}{\widetilde{\Sigma}}. \ee By $m$
we denote the ADM-mass and by $a$ the angular momentum of Kerr
spacetime. To choose a suitable height function for a hyperboloidal
time coordinate in the exterior domain, we use the Taylor expansion of
(\ref{eq:height_asym}). It reads
\[   h'(\tr) = 1+\frac{2m}{\tr} +
\frac{8m^2-C-a^2\sin^2\vartheta}{2 \tr^2} +
O\left(\frac{1}{\tr^3}\right), \quad \mathrm{as} \ \tr\to\infty.\] We
use the first two terms and set $h(\tr) = \tr + 2m \ln \tr$, as for
Schwarzschild spacetime in Schwarzschild coordinates. It is
interesting that the height function for a hyperboloidal foliation
does not necessarily depend on angular momentum but on mass. On the
other hand, the form of the height function should not surprise us, as
asymptotically, the Kerr metric in Boyer-Lindquist coordinates
approaches the Schwarzschild metric in Schwarzschild coordinates. With
the above choice of the height function, the components of the metric
in the exterior domain of Kerr spacetime in a scri-fixing gauge take
the form
\begin{eqnarray} \label{eq:kerr_bl}
g_{tt} &=& -\left(1-\frac{2mr\Omega}{\Sigma}\right)\Omega^2, \quad g_{tr} =
- \left(1-\frac{2mr\Omega}{\Sigma}\right)
\left(1+\frac{2m\Omega}{r}\right),\nonumber \\
g_{t\varphi} &=&  - \frac{2 a m r \Omega^3}{\Sigma} \sin^2\vartheta, \quad
g_{r\varphi} = -\frac{2 a m \Omega}{\Sigma} (r+2m\Omega)\sin^2\vartheta,  \\
g_{rr} &=& \frac{1}{\Omega^2}\left(\left(\frac{2 m \Omega r}{\Sigma}-1\right)
  \left(\frac{2 m \Omega}{r}+1\right)^2+\frac{\Sigma }{\triangle
   }\right), \quad
g_{\vartheta\vartheta} = \Sigma, \quad g_{\varphi\varphi}=
R^2\sin^2\vartheta,\nonumber
\end{eqnarray}
where we have set $\Sigma = \Omega^2 \widetilde{\Sigma}$,
$\triangle=\Omega^2\widetilde{\triangle}$, and $R = \Omega
\tilde{R}$. A calculation shows that the rescaled metric component
$g_{rr}$ is manifestly regular at $\{\Omega=0\}$ and can be written as
\[ g_{rr} = \frac{8m^2}{\Sigma}\left(1+\frac{m\Omega}{r}\right) -
\frac{a^2\sin^2\vartheta}{\triangle} -\frac{4m^2}{r^2} +
\frac{2m}{r\triangle\Sigma} \left(r^2(2mr-a^2\Omega) -
a^2\Omega\cos^2\vartheta(2a^2\Omega^2+r(r-4m\Omega))\right). \] 
The rescaled metric becomes at future null infinity
\[ g|_{\scri^+} =  - 2\,dt\,dr+(8m^2-a^2\sin^2\vartheta)\,dr^2 +
d\vartheta^2 + \sin^2\vartheta\,d\varphi^2. \] We see that, in these
coordinates, the coordinate speed of outgoing characteristics depends
on $\vartheta$ in a manner related to the angular momentum of the Kerr
black hole.
\subsection{In Kerr-Schild coordinates}
The Kerr metric in ingoing Kerr-Schild coordinates reads
\begin{eqnarray*} \tilde{g} &=&
  -\left(1-\frac{2m\tr}{\widetilde{\Sigma}}\right) d\tt^2
  +\frac{4m\tr}{\widetilde{\Sigma}}\,d\tt d\tr - \frac{4 a m
    \tr}{\widetilde{\Sigma}}\sin^2\!\vartheta \,d\tt\,d\varphi -
  2a\sin^2\!\vartheta\left(1+\frac{2m\tr}{\widetilde{\Sigma}}\right)
  \,d\tr d\varphi+\\ && +
  \left(1+\frac{2m\tr}{\widetilde{\Sigma}}\right)\,d\tr^2 +
  \widetilde{\Sigma}\,d\vartheta^2 +
  \tilde{R}^2\sin^2\vartheta\,d\varphi^2, \end{eqnarray*} with the
definitions (\ref{eq:kerrdef}). For the choice of the height function
we use the Taylor expansion of (\ref{eq:height_asym}) as before and
get
\[  h'(\tr) = 1+\frac{4m}{\tr} +
\frac{16m^2-C}{2 \tr^2} + O\left(\frac{1}{\tr^3}\right), \quad
\mathrm{as} \ \tr\to\infty.\] Integrating the first two terms gives
$h(\tr) = \tr + 4m \ln \tr$. With this choice of the height function,
the components of the resulting metric in the exterior domain of Kerr
spacetime in a scri-fixing gauge take the form
\begin{eqnarray} \label{eq:kerr_ks}
g_{tt} &=& -\left(1-\frac{2mr\Omega}{\Sigma}\right)\Omega^2, \quad
g_{tr} = -\left(1+\frac{4m\Omega}{r}\right) +
\frac{4m\Omega}{\Sigma}(r+2m\Omega),\nonumber \\ g_{t\varphi} &=& -
\frac{2 a m r \Omega^3}{\Sigma} \sin^2\vartheta, \quad g_{r\varphi} =
-a\sin^2\vartheta \left(1+\frac{4m\Omega}{\Sigma}(r+2m\Omega)\right),
\\ g_{rr} &=& \frac{8m}{r^2\Sigma}(r+2m\Omega)(2mr-\Omega
a^2\cos^2\!\vartheta), \quad g_{\vartheta\vartheta} = \Sigma, \quad
g_{\varphi\varphi}= R^2\sin^2\vartheta,\nonumber
\end{eqnarray}
The rescaled metric becomes at future null infinity
\[ g|_{\scri^+} =  - 2\,dt\,dr+16m^2\,dr^2 -
2a\sin^2\!\vartheta\, dr\, d\varphi + d\vartheta^2 +
\sin^2\vartheta\,d\varphi^2. \] 
\section{Discussion}\label{sec:conc}
It has long been suggested that using a hyperboloidal foliation in
combination with conformal compactification of the metric to include
null infinity in the computational domain would be a solution to the
outer boundary problem as well as the radiation extraction problem
within the 3+1 approach in numerical relativity \cite{Friedrich83a,
  Huebner98}. This method requires an adequate choice of coordinate
and conformal gauge in the asymptotic domain. An especially suitable
class of gauges for numerical studies on radiative properties of
fields is scri-fixing \cite{Frauendiener98b,Husa05}. While it has been
known how to construct an explicit scri-fixing gauge in Minkowski
spacetime \cite{Moncrief00, Husa02b}, there was no such construction
available for black hole spacetimes. In this article we have
explicitly constructed and studied scri-fixing gauges in Minkowski,
Schwarzschild and Kerr spacetimes. The presented constructions are in
such a generality that the methods can be applied to any explicitly
given asymptotically flat spacetime admitting a smooth conformal
compactification at null infinity.

We may argue that scri-fixing gauges are both natural and useful.
Observers with respect to which the coordinate location of $\scri^+$
is independent of time represent asymptotically stationary
observers. We have seen in the examples that in the asymptotic limit,
Killing observers become null. This is also seen by the fact that a
scri-fixing gauge in a stationary spacetime can be constructed such
that the metric is manifestly time independent. In this sense,
scri-fixing reflects the astrophysical situation that the
gravitational wave detectors are asymptotically stationary observers
of an isolated system. It is also with respect to such observers that
notions of conserved quantities can conveniently be defined. Choosing
a scri-fixing gauge would considerably simplify the calculation of
such quantities and the analysis of numerical solutions. Due to the
manifest regularity of the rescaled metric in our construction, points
at infinity can be treated on the same level as finite points.

A further suitable property of the foliations we discussed is their
intrinsic time asymmetry. Many calculations in numerical relativity
are made with time symmetric data. It seems highly unlikely that such
data can be regarded as physically reasonable. Hyperboloidal surfaces on
the other hand do not allow time symmetry by their very nature and
therefore reflect the physical situation more closely.

That an explicit scri-fixing gauge can be useful has been demonstrated
in \cite{Fodor03, Fodor06} in Minkowski spacetime. A natural next step
would be to extend such studies to black hole spacetimes and to other
test fields. One can also study to what extent the outer boundary
introduced in currently common numerical calculations influences the
solution. In particular, the matching technique would allow us to make such
a comparison conveniently. 

The main motivation to include null infinity in the computational
domain is to have access to the correct asymptotic waveform. It seems,
however, that the approximations commonly used in numerical relativity
may be sufficiently accurate for many purposes. An interesting
question to study in this context would be whether and to what extent
numerical access to null infinity is relevant for astrophysically
realistic predictions concerning gravitational radiation signals to be
measured on Earth \cite{Winicour92}. Just to give an example, it is
known that the tail behavior is different at null infinity and at a
finite distance away from the source \cite{Gundlach93a, Dafermos05,
  Purrer:2004nq}. In this case, a naive interpretation of finite
distance extractions may even lead to a misleading error estimate
\cite{Zenginoglu:2008wc}. Certain other features of gravitational
radiation might also depend on the observers location in a non-trivial
manner.

One may gain some insight from the conformal compactifications
presented in this article for the construction of hyperboloidal
initial data \cite{Andersson02a}. The examples may also serve as a
test ground for new ideas on treating the hyperboloidal initial value
problem for the Einstein equations. The construction of a scri-fixing
gauge in this case can be achieved by prescribing a suitable
representation for the conformal factor in terms of coordinates in
combination with an appropriate choice of a gauge source function in
the general wave reduction of the Einstein equations
\cite{Zeng08b}. Whether this approach can be implemented in numerical
calculations remains to be seen. The explicit examples presented in
this article may be useful for performing numerical experiments in
this approach.

\begin{acknowledgments}
I am grateful to Helmut Friedrich and Sascha Husa for discussions and
comments on the manuscript. I would also like to thank a referee for
constructive suggestions.
\end{acknowledgments}

\bibliography{references}\bibliographystyle{apsrmp}
\end{document}